%% file: iclr2026_conference.tex
\documentclass{article}
\usepackage{iclr2026_conference,times}

\input{math_commands.tex}

\usepackage{hyperref}
\usepackage{url}
\usepackage{graphicx}
\usepackage{algorithm}
\usepackage{algorithmic}
\usepackage{amsmath,amssymb}
\usepackage{multirow}
\usepackage{booktabs}
\usepackage{caption}
\usepackage{makecell}
\usepackage{tcolorbox}
\tcbuselibrary{breakable}
\usepackage{verbatim}
\usepackage{xcolor}
\usepackage{wrapfig}

\title{ACE: Self-Evolving LLM Coding Framework \\ via Adversarial Unit Test Generation and Preference Optimization}

\author{
Yixu Huang \quad
Xinglei Yu \quad
Zhongyu Wei \\
School of Data Science \\
Fudan University \\
Shanghai, China \\
\texttt{\{yixuhuang23, yuxl23\}@m.fudan.edu.cn, zywei@fudan.edu.cn}
}

\iclrfinalcopy 
\begin{document}

\maketitle

\begin{abstract}
Large Language Models (LLMs) excel at code generation but remain heavily reliant on large-scale annotated solutions and verification-based supervision, which constrains scalability and hinders sustained self-improvement. Recent solver-verifier frameworks exploit program execution as an automatic supervision signal, but their effectiveness degrades as solvers become moderately strong: verifier-generated tests increasingly confirm semantic correctness rather than exposing the remaining failure modes.
We propose \textbf{ACE}, a self-evolving code generation framework based on a solver–adversary architecture that prioritizes active failure discovery through execution-centric supervision. A single LLM alternates between generating candidate programs and producing adversarial unit test inputs optimized to induce execution-level failures, such as runtime errors, exceptions, or non-termination. Supervision is derived solely from execution outcomes: robust programs are selected for supervised fine-tuning, while adversarial tests are optimized via Kahneman–Tversky Optimization using execution-derived preferences. Notably, the entire training loop requires no ground-truth code, or external reward models.
Experiments on CodeContests, MBPP, and LiveCodeBench demonstrate that ACE consistently outperforms strong solver-verifier baselines, achieving 3–7\% absolute gains in pass@1, with larger improvements on out-of-distribution benchmarks, while maintaining competitive or improved inference efficiency.
\end{abstract}

\section{Introduction}

LLMs have demonstrated remarkable progress in fields such as mathematical reasoning and code generation, but their performance remains heavily dependent on large-scale, ground-truth training data~\citep{jiang2024survey,huynh2025large}. This heavy dependence incurs high annotation costs and caps achievable performance, spurring increased interest in self-evolving paradigms where models iteratively improve using their own generated data. Recent advances in self-evolution and self-refinement suggest that, under appropriate supervision signals, LLMs can iteratively enhance their capabilities without continuous human intervention~\citep{madaan2023self,tao2024survey,xiong2025self}.

Code generation provides a particularly suitable setting for self-evolving LLMs, as program execution naturally induces an automatic and verifiable supervision signal with clear success or failure outcomes. Building on this property, recent work has explored solver-verifier paradigms for continual improvement in code generation~\citep{liu2024dstc,jin2025reveal,wang2025co}. In this framework, a solver generates candidate programs, while a verifier constructs executable unit tests with expected outputs to assess semantic correctness, and both components are iteratively optimized.

\begin{figure}[t]
\begin{center}
\includegraphics[width=0.98\linewidth]{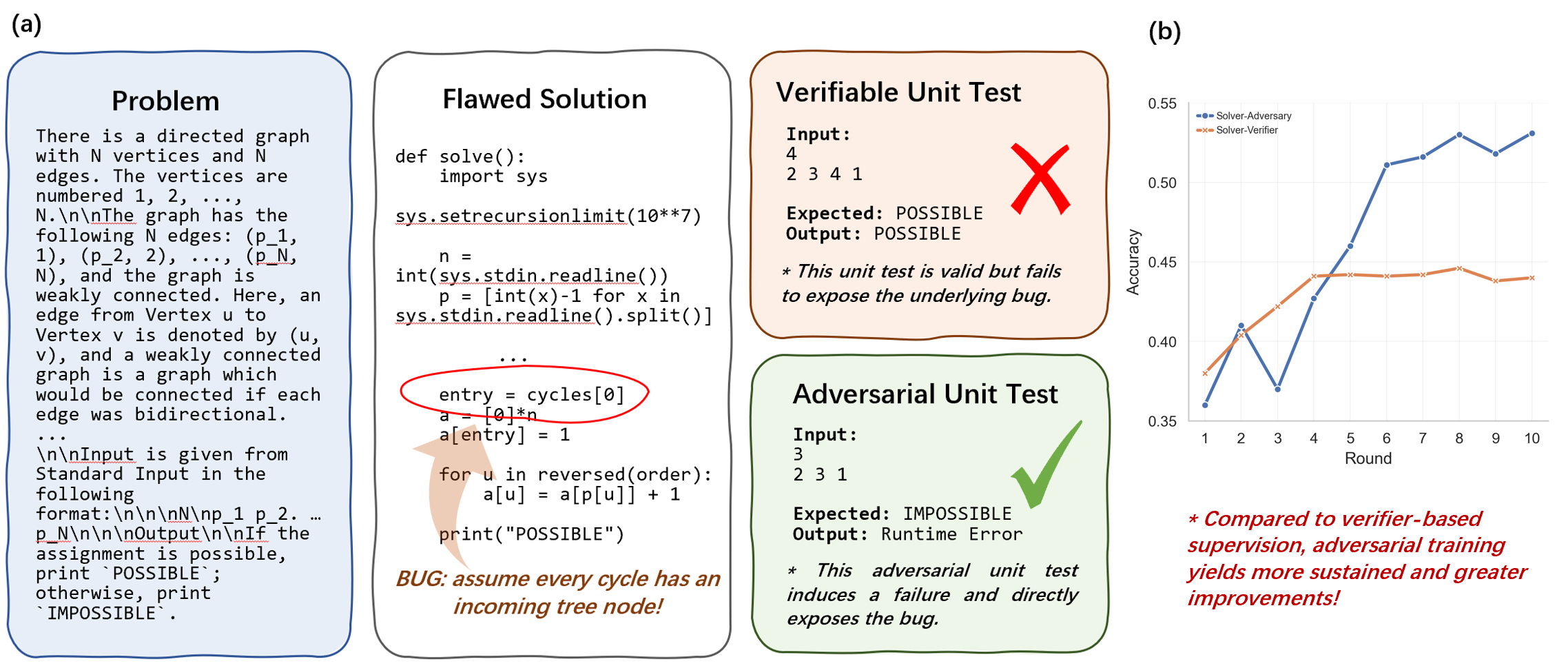}
\end{center}
\caption{
Comparison between verifiable and adversarial unit tests.
\textbf{(a)} An illustrative example where a verifiable unit test passes successfully but fails to reveal a latent bug in the flawed solution. The adversarial unit test, by contrast, induces a runtime error and directly exposes the incorrect assumption that every cycle has an incoming tree node.
\textbf{(b)} Accuracy trends over training rounds under both solver-verifier and solver–adversary structures.
}
\label{fig:verifiable_vs_adversarial}
\end{figure}

However, verifier-based supervision is inherently limited to input spaces where expected outputs can be reliably computed. As the solver attains moderate proficiency, most generated programs pass semantic checks on these verifiable cases, causing verifier-generated tests to quickly saturate and forfeit their discriminative power (see Figure~\ref{fig:verifiable_vs_adversarial}(b)). As a result, the training signal becomes increasingly biased toward already-solved instances, leaving subtler failure modes—such as corner cases, boundary violations, and runtime fragility—largely unaddressed. This saturation stems from a fundamental constraint of semantic verification itself, rather than insufficient verifier capability, as illustrated in Figure~\ref{fig:verifiable_vs_adversarial}(a).   

To address this limitation, we propose \textbf{ACE}, a self-evolving code generation framework based on a solver–adversary architecture. Unlike verifiers that assess semantic correctness by computing expected outputs, the adversary is explicitly optimized to generate adversarial unit test \emph{inputs} that probe execution behavior, using execution-level failure signals (e.g., runtime errors, exceptions, or non-termination) to expose robustness failures in solver-generated code. Inspired by fuzzing and adversarial testing in software engineering~\citep{godefroid2008automated,yaghoubi2019gray}, this execution-centric supervision signal is complementary to verifier-based feedback and remains informative beyond the verifier’s effective coverage.

The framework uses a single LLM that alternates between generating candidate programs as solver and producing challenging test inputs as adversary in a multi-round loop. In each round, adversarial tests are executed on multiple solver outputs. Based on execution outcomes, we distinguish strong adversarial tests that induce failures from weak ones, and select robust programs that pass adversarial scrutiny for training. We jointly optimize the same LLM using these execution-derived signals: the solver via supervised fine-tuning (SFT)~\citep{wei2021finetuned} on filtered high-quality solutions, and the adversary via Kahneman-Tversky Optimization (KTO)~\citep{ethayarajh2024kto} over test preferences. This iterative execution-driven process achieves continual self-improvement without external human supervision, ground-truth code, or additional reward models. Evaluations on three coding benchmarks show that ACE outperforms strong baselines in both code generation performance and robustness, with adversarial tests consistently revealing harder failure modes and enabling sustained gains over existing approaches.

In this work, we make the following key contributions:

\begin{itemize}
    \item We propose ACE, a self-evolving code generation framework based on a solver-adversary architecture. The model alternates between generating programs as solver and producing adversarial unit test inputs as adversary. This introduces an additional supervision channel that emphasizes active failure exposure and robustness.

    \item We show that execution outcomes alone suffice to construct preference signals for both solver and adversary. Adversarial test executions filter robust programs for SFT on the solver and induce desirable-undesirable preference pairs for KTO on the adversary. The entire loop requires no ground-truth code, or external reward models.

    \item Extensive experiments on CodeContests, MBPP, and LiveCodeBench demonstrate that ACE consistently outperforms comparable solver-verifier baselines, with 3-7\% absolute gains in pass@1 and corresponding improvements in pass@5 and pass@10. Gains are largest on out-of-distribution benchmarks, indicating stronger robustness and generalization. Inference efficiency measured by average tokens per solution remains competitive or superior.
\end{itemize}

\section{Related Work}

\paragraph{Recursive Self-Improvement in LLMs.}
Manually annotating large-scale training data is costly, motivating research on improving LLMs through their own generated data.
Early works such as STaR~\citep{zelikman2022star}, Self-Refine~\citep{madaan2023self}, and Self-Play Fine-Tuning~\citep{chen2024self} propose iterative frameworks in which models bootstrap their performance by generating intermediate outputs, feedback, or synthetic training data.
More recent studies extend this paradigm to \emph{coding} LLMs, where execution feedback or program verification provides a natural supervision signal.
Methods such as ReVeal~\citep{jin2025reveal} and CURE~\citep{wang2025co} explore how code generation and validation can be jointly leveraged to form self-improving training loops.

\paragraph{Fuzzing and Adversarial Test Generation.}
Fuzzing and adversarial test generation have long been studied in software engineering as effective techniques for uncovering hidden bugs and corner cases. Prior work including TitanFuzz~\citep{deng2023large} and UAgent~\citep{fu2025uagent} improves code robustness by generating crash-inducing or bug-revealing inputs, while other approaches, such as ATGen~\citep{li2025atgen}, explicitly formulate adversarial test generation as a learning objective and treat test generation as an end goal. 
Despite their success, most adversarial testing methods are primarily studied as standalone robustness tools or test-generation objectives, rather than as components of an end-to-end self-evolving training loop for coding LLMs.
In contrast, many existing self-improvement frameworks rely on validation-oriented testing signals that emphasize correctness confirmation, as exemplified by SelfCodeAlign~\citep{wei2024selfcodealign} and DSTC~\citep{liu2024dstc}.
Our work bridges these directions by embedding adversarial test generation into a multi-round self-evolving framework, where test utility is defined through execution-level discriminative power and directly coupled with model updates.

\paragraph{Preference Optimization for Reinforcement Learning.}
Preference optimization has emerged as a compelling alternative to traditional reinforcement learning approaches that rely on explicit reward modeling.
While early Reinforcement Learning from Human Feedback (RLHF) methods typically adopt policy gradient algorithms such as PPO~\citep{schulman2017proximal}, recent approaches including DPO~\citep{rafailov2023direct} and KTO~\citep{ethayarajh2024kto} reformulate policy optimization as supervised learning over preference signals.
Compared to DPO, which recent frameworks such as DSTC~\citep{liu2024dstc}, SOL-VER~\citep{lin2025learning} utilized, KTO is able to relax the requirement of paired positive–negative samples and supports asymmetric and unbalanced preference data, making it particularly suitable for scenarios where desirable and undesirable outcomes are unevenly distributed. Accordingly, our framework combines KTO with SFT to better accommodate unbalanced preference signals and improve training robustness and efficiency.

\section{Method}

\begin{figure}[h]
\begin{center}
\includegraphics[width=1\linewidth]{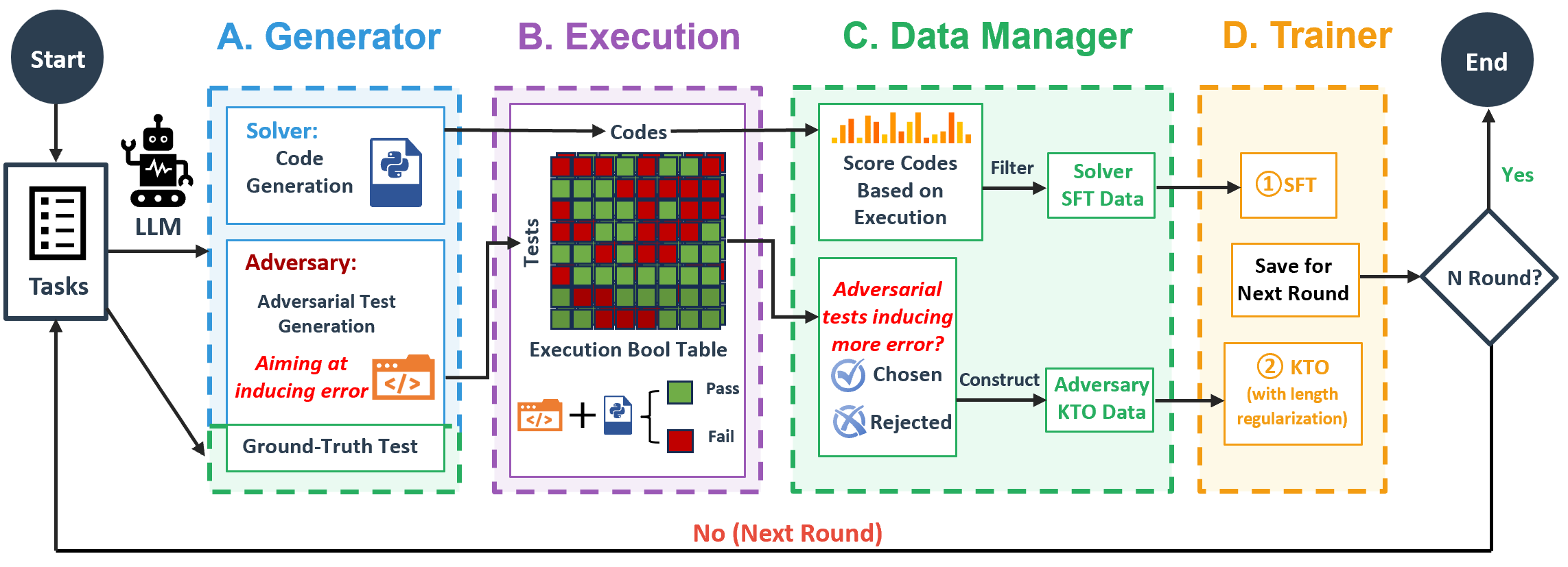}
\end{center}
\caption{Overview of ACE Method. A single LLM alternates between the Solver role, generating candidate codes, and the Adversary role, generating adversarial unit test inputs. Execution outcomes on both ground-truth and adversarial tests are used to construct an execution boolean table, which facilitates code filtering for SFT and preference construction for adversary optimization via KTO. This solver–adversary self-evolution process is repeated for multiple rounds, enabling recursive self-improvement.}
\label{fig:method_pipeline}
\end{figure}

This section introduces the proposed solver–adversary self-evolving framework, as illustrated in Figure~\ref{fig:method_pipeline}. We first describe the adversarial unit test generation process, followed by the construction of SFT data and the SFT procedure. Finally, we show how execution feedback is converted into preference signals to optimize and strengthen the adversary.

\begin{algorithm}[t]
\caption{ACE}
\label{alg:ace}
\begin{algorithmic}[1]
\STATE Initialize shared model parameters $\theta_0$
\FOR{$r = 1$ to $R$}
    \STATE Initialize SFT buffer $\mathcal{B}_{\mathrm{SFT}} \leftarrow \emptyset$
    \STATE Initialize preference buffers $\mathcal{D}_{\mathrm{des}}, \mathcal{D}_{\mathrm{undes}} \leftarrow \emptyset$

    \FOR{each problem $p \in \mathcal{D}$}

        \STATE \textbf{Solver sampling:}
        $
        \mathcal{C}(p) = \{c_i\}_{i=1}^{k_1}
        \sim \pi^{\mathrm{solver}}_{\theta_{r-1}}(\cdot \mid p)
        $

        \STATE \textbf{Adversary sampling:}
        $
        \mathcal{T}^{\mathrm{adv}}(p) = \{t^{\mathrm{adv}}_j\}_{j=1}^{k_2}
        \sim \pi^{\mathrm{adv}}_{\theta_{r-1}}(\cdot \mid p)
        $

        \STATE Execute each $c_i \in \mathcal{C}(p)$ on $\mathcal{T}^{\mathrm{GT}} \cup \mathcal{T}^{\mathrm{adv}}(p)$, recording correctness for ground-truth tests and whether execution results in errors on adversarial tests.

        \STATE Construct execution boolean matrix
        $
        \mathbf{E} \in \{0,1\}^{k_1 \times (|\mathcal{T}^{\mathrm{GT}}| + k_2)}
        $

        \STATE \textbf{Code score:}
        $$
        r_i^{\mathrm{GT}} =
        \frac{1}{|\mathcal{T}^{\mathrm{GT}}|}
        \sum_{t \in \mathcal{T}^{\mathrm{GT}}}
        \mathbb{I}[\mathbf{E}(c_i,t)=1],
        \quad
        r_i^{\mathrm{adv}} =
        \frac{1}{k_2}
        \sum_{j=1}^{k_2}
        \mathbb{I}[\mathbf{E}(c_i,t^{\mathrm{adv}}_j)=1],
        $$
        $$
        s_i = \alpha r_i^{\mathrm{GT}} + (1-\alpha) r_i^{\mathrm{adv}}
        $$

        \STATE \textbf{Solver data selection:} \\
        \[
        \mathcal{C}_{\mathrm{SFT}}(p)
        =
        \mathrm{Top}_{\rho}
        \bigl(
        \{ c_i \in \mathcal{C}(p) \mid s_i \ge \tau \}
        \bigr),
        \mathcal{B}_{\mathrm{SFT}} \leftarrow \mathcal{B}_{\mathrm{SFT}} \cup \{(p, c) \mid c \in \mathcal{C}_{\mathrm{SFT}}(p)\}
        \]

        \STATE  Extract adversarial execution boolean matrix
        $
        \mathbf{E}^{\mathrm{adv}} \in \{0,1\}^{k_1 \times k_2}
        $
        
        \STATE \textbf{Preference data construction:} \\
        \[
        e_j = \sum_{i=1}^{k_1} \mathbb{I}[\mathbf{E}^{\mathrm{adv}}_{i,j}=0], \quad
        \quad
        s_j = k_1 - e_j, \quad
        y_j =
        \mathbb{I}\bigl[
        e_j \ge 1 \;\wedge\; s_j \ge 1
        \bigr]
        \]
        \[
        \mathcal{D}_{\mathrm{des}}
        \leftarrow
        \mathcal{D}_{\mathrm{des}} \cup
        \{(p, t^{\mathrm{adv}}_j) \mid y_j = 1\}, \quad
        \mathcal{D}_{\mathrm{undes}}
        \leftarrow
        \mathcal{D}_{\mathrm{undes}} \cup
        \{(p, t^{\mathrm{adv}}_j) \mid y_j = 0\}
        \]

    \ENDFOR

    \STATE \textbf{Solver update (SFT):}
    $\theta_r \leftarrow
    \mathrm{SFT}(\theta_{r-1}, \mathcal{B}_{\mathrm{SFT}})$
    
    \STATE \textbf{Adversary update (KTO):}
    $\theta_r \leftarrow
    \mathrm{KTO}(\theta_r, \mathcal{D}_{\mathrm{des}}, \mathcal{D}_{\mathrm{undes}})$
    
\ENDFOR

\STATE \textbf{Output:} final model parameters $\theta_R$

\end{algorithmic}
\end{algorithm}

\begin{figure}[t]
    \vspace{-10pt}
    \centering
    \refstepcounter{figure} 
    \begin{minipage}{0.44\linewidth}
        \centering
        \includegraphics[width=\linewidth]{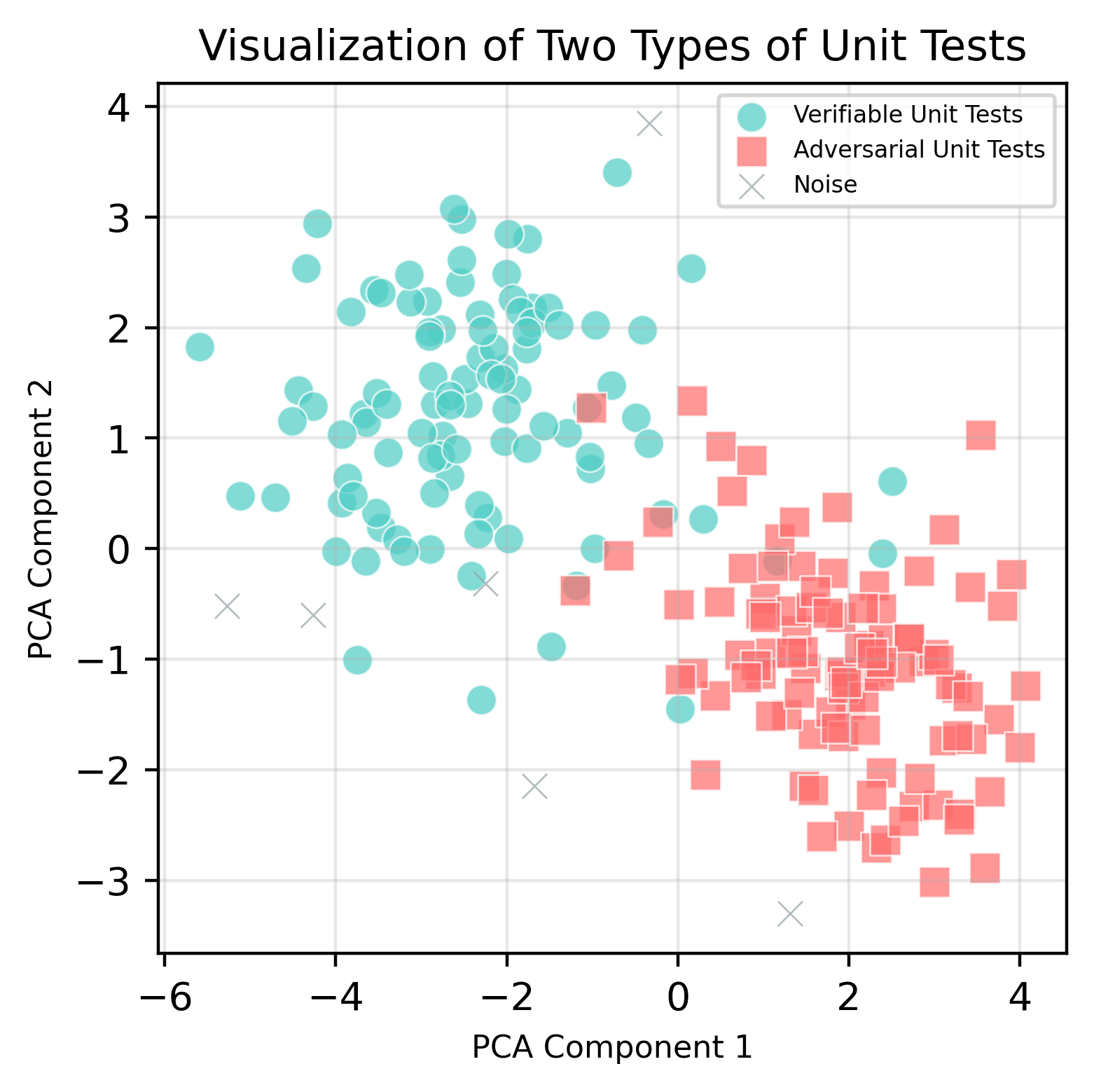}
    \end{minipage}\hfill
    \begin{minipage}{0.50\linewidth}
        \small
        Figure~\thefigure: 
        Visualization of verifiable and adversarial unit tests in a low-dimensional representation space.
        Verifiable tests (blue) and adversarial tests (red) form two visually distinct clusters, indicating a clear distributional separation between the two test types.
        This qualitative difference suggests that adversarial tests occupy a different region of the representation space than verifier-style tests, providing complementary execution signals.
    \end{minipage}
    \label{fig:test_pca}
    \vspace{-10pt}
\end{figure}

\paragraph{Adversarial Unit Test Generation.}
ACE reframes unit test generation as an explicitly adversarial process driven solely by execution behavior, rather than correctness verification against ground-truth outputs.
Given a programming problem, the adversary generates unit test \emph{inputs} without specifying expected outputs.
Each test is executed against a set of solver-generated candidate programs, and only execution outcomes—successful termination or runtime errors—are recorded in an execution boolean table $\mathbf{E}$.
Unlike verifier-style tests that aim to confirm correctness, the adversary is optimized to actively induce execution failures across candidate solutions, thereby exposing latent bugs, unhandled corner cases, and violated assumptions.
Figure~\ref{fig:test_pca} provides an intuitive visualization of this difference, illustrating the representation-level separation between verifiable and adversarial unit tests.
By relying exclusively on execution behavior, adversarial tests offer informative supervision signals for identifying solver weaknesses, without requiring expected outputs to be inferred by model itself.
All executions are conducted in a sandboxed environment with strict time and memory constraints.
Tests that violate the input specifications or exceed resource limits are discarded. In addition, we remove tests that cause all candidate programs to fail, as such cases are observed to reflect invalid inputs rather than providing meaningful adversarial signals.
The remaining execution outcomes are used for solver data selection and adversary preference construction, as described next.
Finally, to provide additional transparency into the nature of adversarial supervision,
we observe characteristics of adversarial tests,
including their types and distribution,
and report the results in Appendix~\ref{sec:adv_test_analysis}.

\paragraph{Code Selection and SFT.}
We construct the solver’s SFT training data by selecting generated codes based solely on execution statistics derived from code--test interactions. For each problem, the solver produces a set of candidate codes $\mathcal{C}=\{c_1,\dots,c_{k_1}\}$, which are executed on both ground-truth tests $\mathcal{T}_{\mathrm{GT}}$ and adversarial tests $\mathcal{T}_{\mathrm{adv}}$. Execution outcomes are recorded in a unified boolean table
\begin{equation}
\mathbf{E} \in \{0,1\}^{k_1 \times (|\mathcal{T}_{\mathrm{GT}}| + |\mathcal{T}_{\mathrm{adv}}|)},
\end{equation}
where $\mathbf{E}_{i,j}=1$ indicates a successful execution under the corresponding test semantics. Specifically, for ground-truth tests,
success requires the program to terminate within resource limits and produce an output that exactly matches the oracle output; for adversarial tests, success only requires normal termination within predefined time
and memory limits. Using this table, we compute the ground-truth pass rate $r_i^{\mathrm{GT}}$ and adversarial execution success rate $r_i^{\mathrm{adv}}$ for each candidate code. We first apply hard filtering by requiring
\begin{equation}
r_i^{\mathrm{GT}} \ge \tau_{\mathrm{GT}}, \qquad r_i^{\mathrm{adv}} \ge \tau_{\mathrm{adv}},
\end{equation}
and compute a combined score
\begin{equation}
s_i = \alpha \cdot r_i^{\mathrm{GT}} + (1-\alpha) \cdot r_i^{\mathrm{adv}},
\end{equation}
and retain only the top fraction $|\mathcal{C}_{\mathrm{SFT}}| \le \rho |\mathcal{C}|$. The selected codes form the SFT dataset used for solver fine-tuning.

\paragraph{Preference Construction and Adversary Optimization.}
For adversary training, we construct preference data based on the same execution table. Given a set of candidate codes and the generated adversarial tests, we count the number of successful and failing executions for each test. A test is labeled as \emph{desirable} if it induces both successes and failures across candidate codes, under valid inputs and within resource limits, indicating that it can effectively discriminate code quality. In contrast, a test is labeled as \emph{undesirable} if all codes succeed as such tests provide no adversarial supervision signal. Notably, a test is discarded if all candidate codes fail, as such cases are ambiguous and may possibly correspond to invalid inputs or violations of problem constraints rather than meaningful adversarial signals. These preference labels are deterministic functions of execution outcomes and require no additional annotation.

Let $\pi_\theta$ denote the adversary policy and $\pi_{\mathrm{ref}}$ a fixed reference policy. For a prompt $p$ and a test input $x$, we define the log-probability ratio
\begin{equation}
\Delta_\theta(x) = \log \pi_\theta(x \mid p) - \log \pi_{\mathrm{ref}}(x \mid p).
\end{equation}
Due to the inherently unbalanced preference data by construction, we adopt KTO to asymmetrically increase the likelihood of desirable tests while suppressing undesirable ones. To discourage excessively long test inputs, we introduce a length penalty and modify the log-ratio as
\begin{equation}
\Delta_\theta^{\mathrm{LP}}(x) = \Delta_\theta(x) - \lambda \ell(x),
\end{equation}
where $\ell(x)$ denotes the token length of $x$. The final objective is a weighted combination over desirable and undesirable samples:
\begin{equation}
\mathcal{L}
=
w_{\mathrm{des}}\,\mathbb{E}_{x \sim \mathcal{D}_{\mathrm{des}}}[\mathcal{L}_{\mathrm{des}}(x)]
+
w_{\mathrm{undes}}\,\mathbb{E}_{x \sim \mathcal{D}_{\mathrm{undes}}}[\mathcal{L}_{\mathrm{undes}}(x)].
\label{eq:length_kto}
\end{equation}
As adversarial testing improves over training rounds, the adversary generates increasingly discriminative tests, yielding sharper execution-based preference signals. These signals in turn enable more effective solver selection and fine-tuning in subsequent iterations, as illustrated in Figure~\ref{fig:round_curve}.

\section{Experiment}

\subsection{Experimental Setup}
We implement ACE on two open-source instruction-tuned backbones, \emph{Qwen3-4B-Instruct}~\citep{yang2025qwen3} and \emph{Qwen2.5-7B-Instruct}~\citep{yang2024qwen25}.
A single backbone alternates between solver and adversary roles throughout self-evolution, without introducing additional models.

Training is conducted on the CodeContests dataset~\citep{li2022competition}, which contains approximately 4.5k programming problems.
In each self-evolution round, the solver samples $16$ candidate programs per problem, while the adversary generates $16$ unit test inputs to probe solver behavior. To encourage diverse generations, we adopt stochastic decoding with temperature set to $1.0$.
The solver is optimized via SFT, and the adversary is optimized using KTO with length regularization. Both processes update model parameters through LoRA~\citep{hu2022lora} with separate adapters.
Detailed prompts and training configurations are provided in the appendix.

For analyzing self-evolution dynamics, we conduct experiments on a fixed 10\% subset of the training data, which we find sufficient to reveal stable trends. All benchmark results are obtained by training on the full dataset and reporting performance after five self-evolution rounds.

\textbf{Benchmarks.}
We evaluate ACE on MBPP~\citep{austin2021program}, LiveCodeBench v2~\citep{jain2024livecodebench}, and a held-out subset of CodeContests.

\textbf{Baselines.}
We compare ACE with representative solver--verifier methods, including ReasonFlux, as well as instruction-tuned baselines with comparable model sizes.

\textbf{Metrics.} We report pass@$k$ ($k\in\{1,5,10\}$) to measure functional correctness, along with the average number of generated tokens per solution to evaluate inference efficiency.All benchmark evaluations use temperature=0.8 for pass@k.

All methods are evaluated under identical decoding configurations.

\subsection{Results}

\subsubsection{Self-evolving Trend}
\begin{wrapfigure}{r}{0.45\linewidth}
    \centering
    \includegraphics[width=\linewidth]{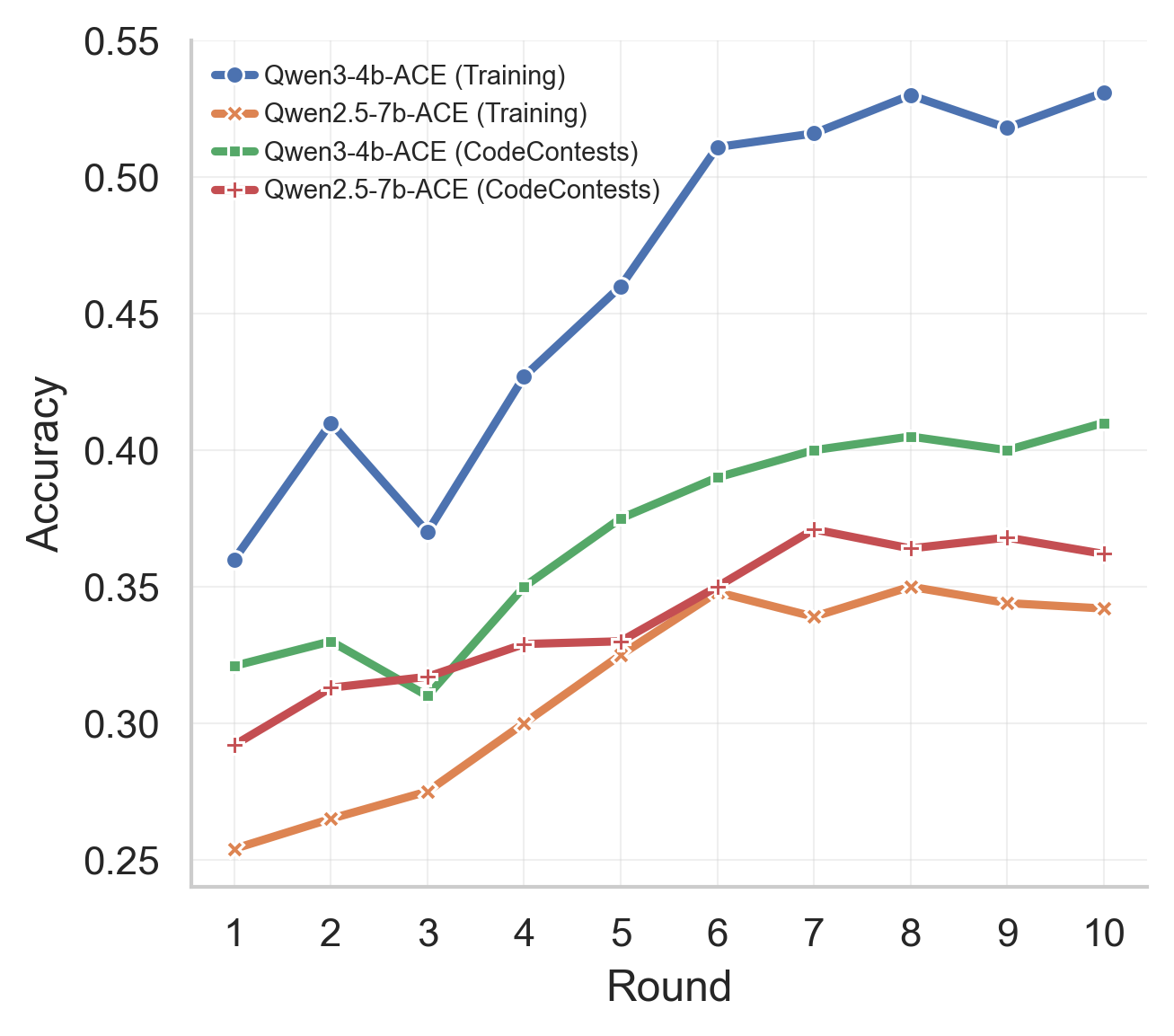}
    \caption{Accuracy over ACE training rounds on the training set and the CodeContests benchmark.
    Both Qwen3-4B-ACE and Qwen2.5-7B-ACE exhibit consistent performance improvements as training progresses and converge after several rounds.}
    \label{fig:round_curve}
\end{wrapfigure}

Figure~\ref{fig:round_curve} shows the evolution of coding accuracy over successive self-evolution rounds.
For both 4B and 7B backbones, ACE improves monotonically in the first several rounds and converges after approximately four to five iterations.
This indicates that adversarial unit tests provide increasingly informative execution feedback, enabling the solver to correct failure cases that were not exposed in earlier rounds.
After convergence, additional rounds yield marginal gains, suggesting that the solver and adversary reach a stable interaction state.

\subsubsection{Benchmark Results}
Table~\ref{tab:main_results} reports results on CodeContests, MBPP, and LiveCodeBench.
CodeContests is treated as an ID benchmark, while MBPP and LiveCodeBench are treated as OOD benchmarks due to differences in problem structure and test design.

\begin{table*}[t]
\centering
\captionsetup[table]{skip=6pt}
\caption{Performance comparison on MBPP, CodeContests, and LiveCodeBench.
We report pass@$k$ and average token consumption (Avg.\ Tok.)
 per benchmark.}
\label{tab:main_results}
\setlength{\tabcolsep}{2.0pt}
\begin{tabular}{l|cccc|cccc|cccc}
\toprule
\multirow{2}{*}{Model} 
& \multicolumn{4}{c|}{MBPP} 
& \multicolumn{4}{c|}{CodeContests} 
& \multicolumn{4}{c}{LiveCodeBench} \\
& @1 & @5 & @10 & \makecell{Avg.\\ Tok.}
& @1 & @5 & @10 & \makecell{Avg.\\ Tok.}
& @1 & @5 & @10 & \makecell{Avg.\\ Tok.} \\
\midrule
Qwen3-4B-Instruct 
& 25.6 & 36.5 & 39.9 & \underline{785.9} 
& \underline{41.7} & \underline{53.3} & \underline{56.5} & \underline{2808.2} 
& 32.4 & \underline{46.9} & \underline{50.6} & 2346.7 \\
ReasonFlux-Coder-4B 
& \underline{33.3} & \underline{46.1} & \underline{50.6} & 2519.2 
& 24.0 & 30.0 & 31.4 & 3554.7 
& \underline{33.2} & 40.8 & 43.4 & 3349.8 \\
Qwen2.5-Coder-3B
& 20.4 & 32.1 & 38.2 & 1277.1 
& 12.7 & 25.4 & 29.7 & \textbf{859.1} 
& 11.6 & 27.8 & 36.0 & \textbf{708.4} \\
Qwen3-4B-\textbf{ACE} 
& \textbf{35.8} & \textbf{49.9} & \textbf{52.3} & \textbf{688.8}
& \textbf{46.7} & \textbf{55.7} & \textbf{60.1} & 3174.3
& \textbf{37.5} & \textbf{49.5} & \textbf{53.4} & \underline{2286.6} \\
\midrule
Qwen2.5-7B-Instruct 
& 24.1 & 33.5 & 39.8 & 430.8 
& 23.2 & 35.6 & 40.2 & \textbf{630.9} 
& 30.4 & 37.2 & 43.7 & \textbf{522.8} \\
ReasonFlux-Coder-7B 
& \textbf{29.5} & \underline{44.8} & \underline{51.7} & \textbf{420.7}
& \underline{26.4} & \underline{38.6} & \underline{43.1} & \underline{641.3} 
& \underline{33.5} & \underline{43.2} & \underline{47.2} & 1563.4 \\
Qwen2.5-Coder-7B
& 23.8 & 35.7 & 48.0 & 2139.2
& 5.0 & 14.5 & 19.7 & 2738.3 
& 4.7 & 17.5 & 26.6 & 2743.2 \\
Qwen2.5-7B-\textbf{ACE} 
& \underline{28.8} & \textbf{46.3} & \textbf{54.2} & \underline{558.8}
& \textbf{29.3} & \textbf{38.8} & \textbf{43.7} & 1002.4
& \textbf{38.9} & \textbf{45.6} & \textbf{49.9} & \underline{1477.8} \\
\bottomrule
\end{tabular}
\end{table*}

\paragraph{In-distribution Performance.}
On CodeContests, ACE achieves the best performance across all pass@$k$ metrics and both model scales.
With the 4B backbone, pass@1 increases from 41.7 for instruction tuning to 46.7 with ACE, while pass@10 improves from 56.5 to 60.1.
The 7B model shows the same trend, where ACE outperforms ReasonFlux on all pass@$k$ metrics.
These results show that adversarial unit test supervision remains effective even when training and evaluation distributions are aligned.

\paragraph{Out-of-distribution Generalization.}
ACE shows larger gains on OOD benchmarks.
On MBPP, ACE improves pass@1 by 2 to 4 points over solver-verifier baselines for both 4B and 7B models, with larger improvements at higher $k$.
On LiveCodeBench, ACE achieves the highest accuracy among all compared methods across all pass@$k$ metrics.
This indicates that adversarial unit tests expose failure modes, leading to strong robustness under distribution shift.

\paragraph{Efficiency and Robustness.}
ACE maintains competitive inference efficiency while improving accuracy.
On MBPP with the 4B model, average token usage decreases from 785.9 to 688.8 while pass@$k$ improves.
Similar trends are observed on other benchmarks, suggesting that adversarial self-evolution encourages more concise solutions rather than longer exploratory generations.
Overall, ACE improves accuracy, generalization, and efficiency simultaneously, demonstrating that solver adversary self-evolution is more effective than solver-verifier training.

\subsection{Ablation Study}
We conduct ablation experiments to examine the contribution of key components in ACE.
All variants are trained with the same backbone, data, and compute budget as the full model.
Quantitative results are reported in Appendix~\ref{sec:ablation_appendix}, and we summarize the main findings below.

\textbf{Adversarial vs.\ Non-Adversarial Tests.}
We replace the optimized adversary with a random test generator that produces unit tests via prompt perturbations, without any adversarial objective or preference optimization.
Although the number of generated tests remains unchanged, this variant leads to a clear performance drop on both the training set and CodeContests (around $4\%$ absolute in pass@1).
This suggests that the gains of ACE do not arise from increased test diversity alone, but from purposefully optimized adversarial tests.

\textbf{Effect of Preference Optimization (No KTO).}
We ablate preference optimization by disabling KTO and using the adversary solely as a test generation module to construct SFT data.
While this variant still benefits from adversarially generated tests, performance degrades substantially compared to the full ACE model, with an absolute drop of over $5\%$ on CodeContests.
This indicates that preference-based optimization is crucial for improving the quality of adversarial tests beyond what can be achieved by supervised learning alone.

\textbf{Effect of Supervised Fine-Tuning (No SFT).}
We further remove the SFT stage and update the solver only via KTO using adversarial execution feedback.
This variant exhibits the largest performance degradation, especially on CodeContests, highlighting that preference-based optimization alone is insufficient for stable solver improvement.
These results demonstrate that SFT and KTO play complementary roles. SFT consolidates reliable behaviors, while KTO emphasizes robustness against adversarial failure cases.

Overall, removing any core component consistently degrades performance.
These ablations confirm that ACE’s improvements stem from the interaction between adversarial test generation, execution-based supervision, and preference-based optimization, rather than from any single component in isolation.

\section{Conclusion and Limitation}
Our experiments demonstrate that adversarial unit test generation yields a more informative execution supervision signal than verifier-style semantic correctness confirmation, particularly when the solver has already achieved high accuracy on verifiable cases. By actively inducing execution failures and exposing solver blind spots that remain invisible to output-based verification, ACE encourages the model to confront challenging edge cases and structural fragilities. This shift in supervision enables more effective self-improvement in coding performance, and empirically leads to more concise and reliable code generation with competitive or reduced inference cost.

In the current implementation, we adopt SFT for solver optimization and KTO for adversary training to prioritize training stability under sparse and noisy execution feedback.
Compared to fully on-policy reinforcement learning, this design avoids high-variance reward signals and enables scalable self-evolution across multiple rounds with consistent performance gains.

Nevertheless, SFT+KTO should be viewed as a pragmatic starting point rather than a fundamental limitation.
The ACE framework is compatible with a broad range of reinforcement learning objectives, including policy gradient and alternative preference-based methods.
Exploring richer RL formulations may further improve exploration efficiency, adversarial diversity, and long-horizon credit assignment, which we leave to future work.

Finally, adversarial test generation introduces potential reward hacking risks, such as learning defensive \texttt{try/except} patterns that mask failures.
Addressing these behaviors through stronger test constraints or regularization remains an important direction for future investigation.

\bibliography{iclr2026_conference}
\bibliographystyle{iclr2026_conference}

\appendix
\section{Appendix}

\subsection{Training Detail}
\label{sec:training_config}

We adopt parameter-efficient fine-tuning for both solver and adversary optimization using LoRA.
Unless otherwise specified, all hyperparameters are fixed across experiments.

\paragraph{LoRA Setup.}
We apply LoRA to the backbone model with rank $r=64$ for all fine-tuning stages.
We use separate LoRA adapters for the solver and adversary on top of a shared backbone.
Adapters are reused across self-evolution rounds within each role.
This design prevents role interference during joint training, while allowing both roles to benefit from shared low-level representations in the backbone.
No quantization is used during training.
No quantization is used during training.

\paragraph{Solver Optimization (SFT).}
Solver fine-tuning is performed using SFT for two epochs per self-evolution round, with a batch size of $4$.
Candidate solutions are filtered based on execution outcomes, requiring a ground-truth test pass rate of at least $0.8$ and an adversarial execution success rate of at least $0.3$.
At most $12.5\%$ of candidate solutions are selected per problem, ranked by a weighted score combining ground-truth and adversarial execution statistics with weight $\alpha=0.6$.

\paragraph{Adversary Optimization (KTO).}
The adversary is optimized using Kahneman--Tversky Optimization (KTO) for two epochs per round with a batch size of $4$.
We apply a length regularization with coefficient $\lambda=0.001$ to discourage overly long test inputs.
Desirable and undesirable samples are weighted equally during preference optimization.

\paragraph{Hardware.}
All experiments are conducted using two NVIDIA RTX PRO 6000 GPUs.

\subsection{Prompt Design}
\label{sec:prompt_details}

To explicitly enforce role switching between the solver and the adversary during self-evolution,
we adopt tailored prompts with clearly specified roles and objectives.
Each role is associated with a fixed prompt template, and only the problem content is instantiated at runtime.
This design ensures a clean separation between solution generation and adversarial unit test construction,
while avoiding prompt-level tuning for individual tasks.

\paragraph{Adversary Prompt.}
The adversary is instructed to generate challenging and discriminative test inputs that expose common implementation errors.
The prompt explicitly encourages reasoning about potential failure modes before producing the final test input.

\begin{tcolorbox}[breakable,colback=gray!5,colframe=gray!40,title=Adversary Prompt Template]
\texttt{<|im\_start|>system} \\
You are a helpful assistant that helps users generate challenging test inputs for coding tasks. \\
\texttt{<|im\_end|>} \\[0.5em]

\texttt{<|im\_start|>user} \\
Given a coding task, your goal is to generate a challenging test INPUT that can expose bugs and weaknesses in code implementations. \\
This is the problem: \\
\{problem\} \\

\{example\_intro\}Your test input should be ADVERSARIAL and CHALLENGING. A good test input should:
\begin{itemize}
    \item Be completely accurate and conform to the problem's input format requirements
    \item Have strong discriminative power to distinguish correct code from buggy code
    \item Target common coding mistakes such as:
    \begin{itemize}
        \item Edge cases (e.g., empty input, single element, extreme values)
        \item Boundary conditions (e.g., off-by-one errors, array bounds)
        \item Special cases (e.g., zeros, negatives, duplicates, overflow)
        \item Corner cases (e.g., identical elements, reverse order, unsorted data)
        \item stress tests (Tests with large input sizes)
    \end{itemize}
\end{itemize}

Before providing a test input, you must think carefully and reason step by step:
\begin{itemize}
    \item What common mistakes might programmers make for this problem?
    \item What edge cases or corner cases are likely to be overlooked?
    \item How can the input expose these mistakes?
\end{itemize}

Finally, after completing the above reasoning steps, output the result in the following format:

\textbf{Test Input:}
\begin{verbatim}
<test input here>
\end{verbatim}

\textbf{Explanation:}
\texttt{<}brief explanation here\texttt{>}. \\
\texttt{<|im\_end|>}
\end{tcolorbox}

\paragraph{Solver Prompt.}
The solver is instructed to generate a correct and executable solution based solely on the problem description and input,
without access to adversarial reasoning or test-generation instructions.

\begin{tcolorbox}[breakable,colback=gray!5,colframe=gray!40,title=Solver Prompt Template]
\texttt{<|im\_start|>system} \\
You are a helpful assistant that helps users solve programming problems. \\
\texttt{<|im\_end|>} \\[0.5em]

\texttt{<|im\_start|>user} \\
You need to think first and then write a Python script.
Your program should use \texttt{input()} to read from standard input and \texttt{print()} to write to standard output.
The output must be computed based on the given input, rather than reproducing any provided examples.

This is the problem: \\
\{problem\} \\
\texttt{<|im\_end|>}
\end{tcolorbox}

We keep the prompt templates fixed across all experiments to ensure consistency and reproducibility.
Role switching is implemented solely by alternating between the solver and adversary prompts,
without modifying decoding strategies or introducing additional heuristics.

\subsection{Analysis of Adversarial Test Types}
\label{sec:adv_test_analysis}
To better understand the nature of adversarial unit tests generated by ACE, we conduct a post-hoc analysis of their observable characteristics. Rather than assigning semantic interpretations to latent representations, we categorize adversarial tests based on surface-level properties and execution behaviors that can be directly measured. This analysis aims to provide qualitative insights into the diversity of adversarial tests, without making assumptions about their underlying semantics.

We group adversarial unit tests into the following categories based on simple and reproducible heuristics:

\begin{itemize}
    \item \textbf{Boundary and extreme-value tests.}
    Tests containing values near the problem-specified limits,
    such as empty inputs, single-element cases, or numerically extreme values.

    \item \textbf{Format-sensitive tests.}
    Tests that stress input formatting, including irregular spacing,
    edge-case line structures, or atypical but valid input layouts.

    \item \textbf{Combinatorial corner cases.}
    Tests that combine multiple challenging conditions simultaneously
    (e.g., extreme values together with degenerate structure),
    which are often difficult to enumerate exhaustively.

    \item \textbf{Large-scale or stress tests.}
    Tests with input sizes significantly larger than the average training examples,
    while still remaining within the specified resource constraints.

    \item \textbf{Other or uncategorized.}
    Tests that do not clearly fall into the above categories.
\end{itemize}

All categorizations are determined using deterministic rules based on input statistics
(e.g., length, value ranges, and structural patterns),
without manual annotation or semantic inspection.

We analyze 100 random adversarial tests collected from the final self-evolution round on the CodeContests training set. For each test, we assign a single dominant category according to the rules above. Table~\ref{tab:adv_test_types} reports the proportion of adversarial tests falling into each category.

\begin{table}[h]
\centering
\caption{Distribution of adversarial unit test types generated by ACE.}
\label{tab:adv_test_types}
\begin{tabular}{l c}
\toprule
\textbf{Test Category} & \textbf{Proportion (\%)} \\
\midrule
Boundary / extreme-value tests & 30.2 \\
Format-sensitive tests & 24.1 \\
Combinatorial corner cases & 22.7 \\
Large-scale or stress tests & 12.3 \\
Other / uncategorized & 10.7 \\
\bottomrule
\end{tabular}
\end{table}

We observe that adversarial tests span multiple categories.
This suggests that the adversary does not collapse to a narrow failure pattern, but is able to explores a diverse set of execution-challenging inputs.

\subsection{Ablation Study Results}
\label{sec:ablation_appendix}

\begin{table}[!htbp]
\centering
\caption{Ablation study of ACE components.
All variants are trained on the full training set using the Qwen3-4B-Instruct backbone,
and evaluated using the checkpoint from round 5.
We report pass@1 on the training set and CodeContests.}
\label{tab:ablation_results}
\setlength{\tabcolsep}{6pt}
\begin{tabular}{lcc}
\toprule
\textbf{Variant} & \textbf{Training} & \textbf{CodeContests} \\
\midrule
Full ACE (Normal) & 46.2 & 34.8 \\
\midrule
Non-adversarial tests (Random) & 41.8 & 30.7 \\
No KTO & 40.8 & 24.9 \\
No SFT & 41.2 & 22.9 \\
\bottomrule
\end{tabular}
\end{table}

\noindent\textbf{Discussion.}
Table~\ref{tab:ablation_results} examines the contribution of key components in ACE.
Removing any core component consistently degrades performance, confirming that ACE’s gains arise from their interaction rather than isolated effects.
Replacing adversarial tests with randomly generated ones leads to a substantial drop in performance, highlighting the importance of actively optimized tests in exposing solver weaknesses beyond surface-level patterns.
Disabling KTO also causes large degradation, suggesting that continuously adapting test generation to the solver’s evolving failure modes is critical for sustained improvement rather than one-shot robustness gains.
Finally, removing SFT results in noticeably worse generalization on CodeContests despite comparable training-set performance, indicating that SFT plays a stabilizing role by consolidating execution-based feedback into transferable behavioral improvements.
Together, these results demonstrate that adversarial testing, preference optimization, and supervised consolidation are complementary components, each of which is necessary for achieving both robustness and generalization in ACE.

\end{document}

%% file: math_commands.tex
\usepackage{amsmath,amsfonts,bm}

\def\eqref#1{equation~\ref{#1}}

\def\1{\bm{1}}

\DeclareMathAlphabet{\mathsfit}{\encodingdefault}{\sfdefault}{m}{sl}
\SetMathAlphabet{\mathsfit}{bold}{\encodingdefault}{\sfdefault}{bx}{n}

